\documentclass[11pt]{article}
\newcommand{\Comment}[1]{{}}
\usepackage[textwidth = 430 pt, textheight = 630 pt]{geometry}
\usepackage{amssymb,euscript,amsmath,amsfonts,graphicx,bbm,simplewick}

\usepackage{color}
\usepackage{bbold}
\definecolor{MyDarkBlue}{rgb}{0.15,0.15,0.45}
\usepackage[linktocpage=true]{hyperref}
\hypersetup{
colorlinks=true,
citecolor=MyDarkBlue,
linkcolor=MyDarkBlue,
urlcolor=MyDarkBlue,
pdfauthor={Constantinos Papageorgakis and Andrew~B.~ Royston},
pdftitle={Revisiting Soliton Contributions to Perturbative Amplitudes},
pdfsubject={hep-th}
}

\usepackage[numbers,sort&compress]{natbib}
\usepackage{hypernat}
\usepackage{slashed}

\newcommand\ignore[1]{}
\def\half{\frac{1}{2}}

\def\tchi{\tilde{\chi}}

\def\bX{\mathbf{X}}
\def\bx{\mathbf{x}}

\def\bk{\mathbf{k}}
\def\bP{\mathbf{P}}

\def\ed{\, \textrm{d}}

\def\one{{\,\hbox{1\kern-.8mm l}}}

 \newcommand{\SU}{\mathrm{SU}}
 \newcommand{\pd}{\partial}

\newcommand{\Cset}{{\,\,{{{^{_{\pmb{\mid}}}}\kern-.45em{\mathrm C}}}}}
 \newcommand{\cA}{\mathcal A}      \newcommand{\cM}{\mathcal M}       
 \newcommand{\nn}{\nonumber}
\newcommand{\ie}{{\it i.e.~}} \newcommand{\eg}{{\it e.g.~}}
 
\newcommand{\be}{\begin{equation}} \newcommand{\ee}{\end{equation}}
\newcommand{\bea}{\begin{eqnarray}} \newcommand{\eea}{\end{eqnarray}}

\newcommand{\FF}{\mathcal{F}} \newcommand{\OO}{\mathcal{O}}
\newcommand{\MM}{\mathcal{M}}

\begin{document}

 \rightline{RUNHETC-2014-03}
 \rightline{QMUL-PH-14-06}
 \rightline{MIFPA-14-09}

   \vspace{2truecm}

\centerline{\LARGE \bf {\sc Revisiting Soliton Contributions}} 
\vspace{.5cm}\centerline{\LARGE \bf {\sc to Perturbative Amplitudes}} 
\vspace{1.5truecm}
\centerline{ {\large {\bf{\sc Constantinos
        Papageorgakis}${}^{\,a,b,}$}}\footnote{E-mail address:
    \href{mailto:Costis Papageorgakis
      <c.papageorgakis@qmul.ac.uk>}{\tt
      c.papageorgakis@qmul.ac.uk}} {and} {\large {\bf{\sc
        Andrew~B.~Royston}${}^{\,c,}$}}\footnote{E-mail address:
    \href{mailto:Andy Royston <aroyston@physics.tamu.edu>}{\tt
      aroyston@physics.tamu.edu} } }

\vspace{1cm}
\centerline{${}^a${\it  NHETC and Department of Physics and Astronomy}}
\centerline{{\it  Rutgers University, Piscataway, NJ 08854-8019, USA}}
\vspace{.5cm}
\centerline{${}^b${\it CRST and School of Physics and Astronomy}}
\centerline{{\it Queen Mary, University of London, E1 4NS, UK}}
\vspace{.5cm}
\centerline{${}^c${\it George~P.~\& Cynthia Woods Mitchell Institute}}
\centerline{{\it for Fundamental Physics and Astronomy}}
\centerline{{\it Texas A\&M University, College Station, TX 77843, USA}}

\vspace{1.5truecm}

\thispagestyle{empty}

\centerline{\sc Abstract}
\vspace{0.4truecm}
\begin{center}
  \begin{minipage}[c]{380pt}{It is often said that soliton
      contributions to perturbative processes in QFT are exponentially
      suppressed by a form factor. We provide a new derivation of this
      form factor for a class of scalar theories with generic soliton moduli. The derivation treats the soliton momenta relativistically and is valid at leading order in momentum transfer. The computation reduces to a matrix element in
      the quantum mechanics on the soliton moduli space. We
      investigate the conditions under which the latter leads to
      suppression. Extending this framework to instanton-solitons in
      five-dimensional Yang-Mills theory leaves open the possibility
      that such contributions will not be suppressed.}
\end{minipage}
\end{center}

\vspace{.4truecm}

\noindent

\vspace{.5cm}

\setcounter{page}{0}

\newpage

\renewcommand{\thefootnote}{\arabic{footnote}}
\setcounter{footnote}{0}

\linespread{1.1}
\parskip 7pt

{}~
{}~

\makeatletter
\@addtoreset{equation}{section}
\makeatother
\renewcommand{\theequation}{\thesection.\arabic{equation}}

\section{Introduction and Summary}

According to standard QFT lore, soliton production is exponentially
suppressed at small coupling and hence unimportant for perturbative
physics. For a theory with a dimensionless effective coupling $g_{\rm
  eff}$ such intuition usually comes from the study of the large-order
behavior of perturbation theory. The basic premise is that the
perturbative expansion is an asymptotic series, which approximates the
full answer for a scattering amplitude up to a remainder term,
\begin{equation}\label{largeorder}
  \mathcal A(g_{\rm eff}) = \sum_{n=0}^{N-1} A_n g_{\rm eff}^{n} +
  \mathcal R_N(g_{\rm eff})\;.    
\end{equation}
The approximation is good when $\mathcal R_N \sim e^{-1/g_{\rm eff}} $ and
this occurs at large orders $N\sim O(1/g_{\rm eff})$, \eg see
\cite{LeGuillou:1990nq}. As such, one would
conclude that at small coupling all contributions which cannot be
accounted for by perturbation theory are exponentially small.

These arguments are most commonly discussed in the context of
instanton contributions to partition functions and correlators.
However, they are also applicable to the process of virtual
soliton-antisoliton pair creation in the following sense
\cite{Drukier:1981fq, Bachas:1992dw}.  If one views the
soliton-antisoliton pair as being composed of a large number, $n\sim
O(1/g_{\rm eff})$, of perturbative particles clustered together at
distances of order one relative to the inverse energy of the incoming
particle, then the contribution of this pair to a perturbative process
would be captured by the remainder function in \eqref{largeorder}
rather than the sum.

There is also a second, different picture for the origin of soliton
suppression,\footnote{See \cite{Banks:2012dp} for a discussion along
  these lines.} which becomes apparent after relating contributions
from virtual soliton pairs to the creation of on-shell
soliton-antisoliton asymptotic states via the optical theorem.  One
can construct a natural parameter, in addition to the coupling, from
the ratio of the soliton Compton wavelength over its size,
$R_C/R_S$. When $R_S\gg R_C$ quantum effects are small and the soliton
can be treated semiclassically. Since this is the regime of
perturbative calculations, it is reasonable to expect that soliton
contributions to a given amplitude are suppressed by factors of
$e^{-R_S/R_C}$.\footnote{In general, $R_C/R_S$ and $g_{\rm eff}$ are
  different parameters. However, note that for the `t
  Hooft--Polyakov monopole in Yang--Mills--Higgs theory $R_C := 1/M
  \propto g_{\rm YM}^2/M_W$, while $R_S\propto 1/M_W$, where $M_W$ is
  the mass of the perturbative $W$-boson. Hence $R_C/R_S\propto g_{\rm
    eff}=g_{\rm YM}^2$. Similar relations hold for the kink in
  two-dimensional $\Phi^4$ theory.}

In line with the soliton literature, we assume that the Compton
wavelength, \ie inverse mass, scales as $g^2$ relative
to a fixed length scale determined from the potential, such as the
inverse mass of a perturbative excitation: $R_C\propto g^2 m^{-1}$.
Thus, {\it if} the soliton size is also fixed in terms of this mass
scale such that $R_S\propto m^{-1}$, then faster-than-any-power
suppression in $R_C/2R_S$ means suppression relative to perturbative
effects. For shorthand we will express functions of this type through
the typical example $e^{-2R_S/R_C}$, although there are of course
other functions with this property. We stress that all such functions
lead to contributions that are suppressed compared to any finite order
in perturbation theory, provided the size $R_S$ is tied to a fixed
scale.

However, our formalism also allows for solitons with moduli-dependent
sizes. Thus a $e^{-2R_S/R_C}$ factor, taken at face value, would imply
finite contributions from configurations with $R_S\propto R_C$. Of
course, when $R_{S} \propto R_C$ one expects that the semiclassical approximation
breaks down. A more conservative stance would
therefore be that small solitons simply invalidate the above arguments
for faster-than-any-power suppression. This still leaves open the
possibility of a drastic modification to the perturbative expansion
and more powerful tools are needed to determine the role of soliton
contributions.

In the main part of this letter we revisit the above discussion by performing a first-principles investigation of soliton contributions in scalar theories, which support solitons with generic moduli spaces.  Our primary assumption will be that crossing symmetry applies to processes involving asymptotic soliton states and moreover that it continues to hold in the semiclassical approximation. This is \eg true for the kink solution in the sine-Gordon model \cite{Coleman:1975ia}.  We begin by employing the optical theorem to associate a soliton loop in a perturbative process with the soliton-antisoliton pair production amplitude. Crossing symmetry then maps the calculation to a form factor in the soliton background, which to leading order in the semiclassical expansion reduces to a matrix element in the quantum mechanics on soliton moduli space. One would then like to compute this matrix element in order to assess exponential suppression.\footnote{The following two paragraphs have been added in 2020 and correct the older version of this paper.\label{footnote}}

In this paper we only utilize the leading term of the collective coordinate quantum mechanics in a time derivative expansion. This implies that the expression we obtain for the form factor is valid only at leading order in the ratio of the momentum transfer to the soliton mass, $k/M_{\rm cl}$. In order to apply crossing symmetry to relate the form factor to the soliton pair production, however, one must have access to the regime of $k \sim O(M_{\rm cl})$, above the pair-production threshold. These points have been addressed in \cite{Melnikov:2020ret}. There it is shown that determining the form factor at $k\sim O(M_{\rm cl})$ requires knowledge of solutions to a wave-like PDE that describes solitons undergoing acceleration.

In light of this, we are not justified in drawing any conclusions regarding the exponential suppression of the pair-production amplitude, based on the large-$k$ behavior of the form-factor obtained here. Nevertheless, our arguments may still be viewed as lending some support to the main message of this paper, which is the following. Even if one could demonstrate exponential suppression in the coupling for the form factor of a soliton of fixed size, it is possible that this suppression will not hold for solitons with a moduli-dependent size that can become small.

We close by sketching an application of this idea to (dyonic) instanton-solitons in 5D Yang--Mills--Higgs theory.  We show how the extra assumption of finiteness, motivated for the maximally supersymmetric theory (MSYM) by its conjectural equivalence to the $(2,0)$ superconformal tensor theory in 6D \cite{Douglas:2010iu,Lambert:2010iw}, provides a self-consistent (and nontrivial) mechanism in which recently obtained perturbative divergences \cite{Bern:2012di} could be canceled by soliton contributions.

\section{Soliton pair production as a form factor}\label{expansion}

Consider the following class of real scalar field theories in
Minkowski space with Lagrangian
\begin{equation}\label{lagrangian}
L = \frac{1}{g^2}\int \ed\bx \left\{ \half \dot{\Phi} \cdot \dot{\Phi}
  -\half\pd_\bx \Phi \cdot \pd_\bx \Phi - V(\Phi) \right\}~.
\end{equation}
We denote by $\bx$ a $(D-1)$-dimensional position vector, while
$\ed\bx$ is shorthand for $\ed^{D-1} x$.  We take the fields to be
$\mathbb{R}^n$-valued, and $\cdot$ denotes the Euclidean dot
product.\footnote{We use italic letters for $D$-dimensional vectors
  and boldface letters for spatial $(D-1)$-dimensional ones.}  Here we
assume that the potential has a dimensionless parameter $g$
controlling the perturbative expansion. Then, in terms of canonically
normalized fields $\tilde{\Phi} = g^{-1} \Phi$, we have
$\tilde{V}(\tilde{\Phi};g) = g^{-2} \tilde{V}(g\tilde{\Phi};1)$, while
we have also set $V(\Phi) = \tilde{V}(g \tilde{\Phi};1)$
\cite{Goldstone:1974gf}.

We are interested in soliton solutions, classically described by
localized, finite-energy field configurations and denoted by
$\phi$. Although Derrick's theorem \cite{Derrick:1964ww,MR2068924}
precludes the existence of soliton solutions in linear sigma models
for $D > 2$, we will for the time being leave $D$ arbitrary. Doing so
will facilitate the extension to theories with gauge interactions
where one can have $D > 2$. It will be clearly indicated in the text
when it becomes necessary to restrict to two dimensions.

For a fixed topological charge sector,\footnote{\label{topocharge}The
  sectors are labeled by homotopy equivalence classes of maps of the
  $D-2$-sphere at infinity into the vacuum manifold, $M_{\rm vac} :=
  \{ \Phi ~|~ V(\Phi) = 0 \}$.} such classical solutions usually come
in a smooth family parameterized by a collection of moduli $U^M$,
where $M = i,m$. In a translation-invariant theory, a subset of these
moduli always consist of the center-of-mass position, $ (U^i) = \bX$.
$U^m$ then parameterize all remaining `centered' moduli. We denote the
moduli space of solutions for a given, fixed topological charge as
$\mathcal M$; it represents a local minimum of the energy functional.
An example of a simple model in two dimensions with a nontrivial
centered moduli space was studied by Rajaraman and Weinberg
\cite{Rajaraman:1975qr}. Our class of two-dimensional models also
contains solitons with moduli-dependent sizes; see \eg
\cite{Bazeia:1997zp,deBrito:2014ega}.

In the presence of a soliton a new sector of the quantum theory opens
up. This is orthogonal to the vacuum sector since solitons carry a
conserved topological charge \cite{Goldstone:1974gf}. Nevertheless,
the soliton-sector single-particle states form a subspace of the total
single-particle Hilbert space and one can study processes involving
both perturbative particles and solitons as asymptotic states. Soliton
states can be chosen to be energy-momentum eigenstates,
$|\bf{P}\rangle$.\footnote{We capitalize the momenta of solitons in
  order to distinguish them from perturbative particle momenta.}  Note
that, in addition to the soliton's momentum, such states can carry
extra labels corresponding to eigenvalues of ancillary operators that
commute with the Hamiltonian. These depend on the particulars of the
theory and will be left implicit for the rest of our discussion.

Let us now study the self-energy of a perturbative
particle, or `meson', of momentum $k$ in the theory \eqref{lagrangian}.  Through
the optical theorem, unitarity of the S-matrix implies that the
imaginary part of any amplitude arises from a sum over a complete
set of intermediate states, \emph{viz}.
\begin{equation}
\label{optical}
2 \textrm{Im} \;{\cal A}({ k}\to { k}) = \sum_f \int \ed \Pi_f |{\cal
  A}({k}\to f )|^2\;.
\end{equation}  
In the above, $|\bf f\rangle$ is a generic multi-particle state of the
theory, the sum is to be taken over the full Fock space and $\Pi_f$ is
the measure for the multi-particle phase space.  In general,
multi-particle states may be constructed from both perturbative and/or
solitonic single-particle states.  However, conservation of
topological charge dictates that only states $|{\bf f}\rangle$ with
zero total topological charge will have nonzero overlap with $|
\bk\rangle$. 

It will be enough for our purposes to concentrate on the simplest such
intermediate configuration consisting of a single soliton-antisoliton
pair of momentum $-P_i$ and $P_f$ respectively and denoted as $|{\bf
f}\rangle = |\bP_f, -\bP_i \rangle$.  We will therefore focus
on the soliton pair-production amplitude:
\begin{align}
  \label{eq:4}
 {\cal A}(k\to P_f,-
  \bar P_i)\;.  
\end{align}
It is unclear how one should proceed with a direct
evaluation of such an expression, since there exists no known
associated analytic classical solution and hence no semiclassical
expansion scheme.\footnote{See however
  \cite{Demidov:2011dk,Demidov:2011eu} for an alternative approach to
  this problem, which would be interesting to compare with the point
  of view taken here.} For that reason, we will employ crossing
symmetry---one of the main axioms in the analytic S-matrix approach to
quantum field theory \cite{MR0204051}---to relate the full amplitude
for pair production to that of a
process where the soliton absorbs the meson:
\begin{align}
  \label{eq:2}
  {\cal A}(k\to P_f,-\bar P_i) = {\cal A}(P_i, k\to P_f)\;.
\end{align}
The advantage of this rewriting is that we can now employ semiclassical
tools to evaluate the expression on the right-hand side. Note that \eqref{eq:2} is
an equality between amplitudes in distinct topological sectors. 

The amplitudes \eqref{eq:2} are nontrivial only when the perturbative
particle is off-shell. Therefore, the right-hand side is related to the form factor
\begin{equation}\label{amplitude}
  i (2\pi)^D \delta^{(D)}(k+P_i-P_f){\cal A}(P_i,k\to P_f)
  = \int \ed^Dx \;e^{-i k\cdot x}\langle {\bf P}_f|{\rm T}\big\{  \Phi(x)  \;e^{-i \int \ed t'
    H_I(t')}\big\} |{\bf P}_i \rangle\;, \quad
\end{equation}
where $H_I$ denotes the interaction Hamiltonian. The Hamiltonian obtained from \eqref{lagrangian} is trivially 
\begin{equation}
H = \int \ed\bx \left\{ \frac{g^2}{2} \Pi \cdot \Pi
  +\frac{1}{ g^2} \Big(\frac{1}{2}\pd_\bx \Phi \cdot \pd_\bx \Phi +
  V(\Phi)\Big)  \right\}\;,
\end{equation}
and its semiclassical expansion in the soliton sector was obtained in
\cite{Papageorgakis:2014jia}. The original conjugate pair $(\Phi,\Pi)$
can be related to the new pairs $(U^M, P_N)$, $(\chi,\pi)$ through the
canonical transformation
\begin{eqnarray}
  \Phi(x) &=& \phi(\bx;U) + g \;\chi(x;U) \cr
  \Pi(x) &=& \half \left( a^M \pd_M \phi(\bx;U) + \pd_M \phi(\bx;U)
    \bar{a}^M \right) + \frac{1}{g} \;\pi(x;U) ~, 
\end{eqnarray}
and subject to the constraints
\begin{equation}\label{constraints}
 F_{1,M}:=\int \ed\bx\; \chi\cdot \pd_M \phi =0\;,\qquad
   F_{2,M}:=\int \ed\bx \;\pi\cdot \pd_M \phi = 0\;,
\end{equation}
which ensure that the fluctuations $\chi,\pi$ are orthogonal to the
zero modes $\pd_M \phi$.  Here we have inserted factors of $g$ so that
the fluctuation fields are canonically normalized.  The functionals
$a^M, \bar{a}^M$ are given by
\begin{equation}\label{theas}
  a^N = \frac{1}{g^2}\left (P_M -\int \pi \cdot \pd_M
    \chi \right) C^{MN}~, \quad \bar{a}^M = \frac{1}{g^2} C^{MN} \left (P_M -
    \int \pd_M \chi  \cdot \pi \right)~, 
\end{equation}
where $C = (G - g \Xi)^{-1}$ with
\begin{equation}\label{xig}
  \Xi_{MN} = \frac{1}{g^2}  \int\chi \cdot\pd_M \pd_N \phi\;,\qquad
  G_{MN} = \frac{1}{g^2}  \int
  \pd_M\phi\cdot \pd_N \phi ~.
\end{equation}
$G_{MN}$ is the metric on moduli space, induced from the flat metric
on field configuration space.

In terms of these new variables the Hamiltonian can be written as
\begin{align}
  \label{eq:3}
H =&~   \frac{g^4}{2}  a^M
 G_{MN}  a^N  + {\rm v}(U^m)  +\int \Big[\half
   \pi \cdot \pi + g\; s\cdot  \chi + 
\frac{1}{2}  \chi \cdot \Delta  \chi + V_I(
\chi)  \Big]  + O(g^2) ~, 
\end{align}
with $V_I(\chi)$ denoting cubic and higher-order interaction terms in
the fluctuations $\chi$ coming from the original potential. In writing
the above, we have ignored operator-ordering ambiguities, such that
$a^M = \bar a^M + O(g^2)$. These corrections correspond to two-loop
effects that will not be important for the rest of our calculation.

We have also defined
\begin{align}\label{sourceOp}
  & s(\bx;U^m) :=\frac{1}{g^2} \Big( - \pd_{\bx}^2 \phi + \frac{\pd V}{\pd \Phi}
  \bigg|_{\Phi = \phi} \Big)~, \qquad
  \Delta  := -\delta_{ab} \pd_{\bx}^2 + \frac{
  \delta^2 V}{\delta \Phi \delta\Phi}\bigg|_{\Phi = \phi}~, \cr 
 & {\rm v}(U^m)  := \frac{1}{g^2} \int \ed\bx \left(
    \half \pd_{\bx} \phi 
    \cdot \pd_{\bx} \phi + V(\phi) \right) = M_{\rm cl} + \delta
  {\rm v}(U^m)\;.
\end{align}
If $\phi$ is an exact solution to the time-independent equations of
motion then $s(\bx;U^m) = 0$ and $\delta {\rm v}(U^m) = 0$.  However
in theories with centered moduli it is sometimes convenient to expand
around a configuration that is only an approximate solution. This will
induce a tadpole for $\chi$ and a moduli-dependent potential.

\section{Evaluation of the nonrelativistic form factor}

We will now use standard techniques to evaluate the form factor
\eqref{amplitude} in the regime of small soliton velocities. The
semiclassical expansion of the Hamiltonian around a slowly-moving
soliton configuration follows from \eqref{eq:3}:
\begin{equation}\label{Hamexp}
  H = H^{(-2)} + H^{(0)} +  O(g)~,
\end{equation}
where 
\begin{eqnarray}
  H^{(-2)} &=&  M_{\rm cl}~, \cr
  H^{(0)} &=& \half P_M G^{MN} P_N +  \delta
  {\rm v}(U) + \half \int \left( \pi \cdot \pi + \chi \cdot \Delta \chi
  \right)~, 
\end{eqnarray}
This expansion is valid
provided we are in the small-velocity and small-(moduli space)
potential approximation:
\begin{align}\label{manton}
  P_M \sim O(1/g)~, \qquad s(\bx;U^m) \sim O(1) \quad \Rightarrow
  \quad M_{\rm cl} \sim O(1/g^2) ~ \textrm{ and } ~ \delta {\rm v}(U^m) \sim
  O(1)~.
\end{align}
These conditions ensure that $\phi(\bx;U^M(t))$ is an approximate
classical solution to the time-dependent equations of motion, in such
a way that the corrections incurred from not expanding around an exact
solution are comparable to the corrections incurred from the
semiclassical saddle-point approximation itself; see
\cite{Papageorgakis:2014jia} for further details.  Given
\eqref{manton} and the fact that $G^{MN}\sim O(g^2)$, one can check
that the $H^{(n)}$ terms in \eqref{Hamexp} are $O(g^n)$. 

It is now easy to see that the leading contributions to the form factor are
\begin{align}\label{leadingpert}
& \int \ed^Dx \;e^{-i k\cdot x}\langle {\bf P}_f| {\rm T}\big\{\Phi(x) e^{-i \int_{-T}^{T}  \ed t' H_I(t')}\big\} | {\bf P}_i\rangle = \cr
& \qquad \qquad \qquad = \int \ed^Dx \;e^{-i k\cdot x} \langle {\bf P}_f| e^{i H (t-T)} \phi(\bx;U) e^{-i H (t + T)}| {\bf P}_i\rangle \left(1 +  O(g)\right) \;.
\end{align}
The latter admits further simplification to leading order where the
dynamics reduce to quantum mechanics on the $d$-dimensional soliton
moduli space $\MM$ \cite{Papageorgakis:2014jia}. This can be
straightforwardly seen by noting that the last two terms in $H^{(0)}$
simply renormalize $M_{\rm cl} + \delta {\rm v}(U)$ to yield
\begin{equation}\label{Hsc}
  \hat H_{\rm s.c.} =  M_{\textrm{1-loop}}+ \frac{1}{2}
    \hat{P}_M \hat G^{MN} \hat{P}_N  
  + \delta \hat {\rm v}_{\textrm{1-loop}} ~.
\end{equation}
In the above we have used $\hat G^{MN} = G^{MN}(\hat{U})$ and placed
hats on $(U^M,P_N)$ to emphasize the fact that they are operators,
satisfying standard commutation relations $[\hat{U}^M,\hat{P}_N] = i
{\delta^M}_N$.

We take a canonical approach to this quantum mechanics where
stationary states are represented by wavefunctions $\Psi(U)$ on moduli
space, such that $(\hat{U}^M \Psi)(U) = U^M \Psi(U)$ and $(\hat{P}_M
\Psi)(U) = -i \pd_M \Psi(U)$.  In particular the state $| \bP \rangle$
may be expanded as
\be\label{Pdef}
|{\bf P}\rangle  = \int \ed^d U\sqrt G \Psi_{{\bf P}}(U)|U\rangle\;.
\ee
Note that $\bP$ is the set of eigenvalues of the center-of-mass
momentum operators, $\hat{P}_i$.  Translational invariance implies
that the full moduli space takes a factorized form,
\begin{equation}
\MM = \mathbb{R}^{D-1}_{\mathbf{X}} \times \tilde{\MM}~,
\end{equation}
with metric
\begin{equation}\label{metfactor}
  \ed s^2 = M_{\rm cl} \ed\bX\cdot\ed\bX + \tilde G_{m 
    n}\ed U^{ m}\ed U^{n} ~.
\end{equation}
The first factor is parameterized by the center-of-mass moduli $(U^i)
= \bX$, while the centered moduli space $\tilde{\MM}$ is
parameterized by the remaining moduli $U^m$.  Additionally,
translational invariance implies that the moduli space potential
$\delta {\rm v}$ is independent of the $U^i$.  It then follows that the
$\hat{P}_i$ commute with the Hamiltonian \eqref{Hsc} and we can
choose our wavefunctions to be simultaneous eigenfunctions of energy
and center-of-mass momentum.  We will denote the energy eigenvalues as
$E_{\bP}$, so that $e^{i \hat{H}_{\rm s.c.} t} \Psi_{\bP} = e^{i E_{\bP} t} \Psi_{\bP}$.  Given \eqref{Hsc} and \eqref{metfactor}, the energy eigenvalues take the form
\begin{equation}\label{nrEvalue}
E_{\bP} = M_{\textrm{1-loop}} + \frac{1}{2 M_{\rm cl}} \bP^2 + \tilde{E}~,
\end{equation}
where $\tilde{E}$ represents the contribution to the energy of the state from the dynamics on the centered moduli space.  Note that the kinetic energy term contributes at the same order, $O(1)$, as the one-loop correction to the rest mass, per \eqref{manton}.  As mentioned at the beginning of our analysis, there might be additional labels
characterizing the soliton state corresponding to extra operators that
commute with the Hamiltonian, in which case they also characterize the
wavefunction.

We can now explicitly write the matrix element appearing in \eqref{leadingpert} as
\begin{align}
& \langle {\bf P}_f| e^{i H (t-T)} \phi(\bx;U) e^{-i H (t + T)}| {\bf P}_i\rangle = \cr
& \qquad \qquad \qquad  = e^{-i (E_i + E_f) T} e^{-i (E_i - E_f )t} \int \ed^d U \sqrt{G} \Psi_f^* \phi(\bx;U)\Psi_i (1+ O(g))\;. \qquad
\end{align}
The factorization of the moduli space \eqref{metfactor} together
with the $\bX$-independence of the potential $\delta {\rm v}(U)$, imply
that the wavefunctions also factorize accordingly:
\begin{equation}\label{wavefunction}
  \Psi_{\bf P}(U^M) = \frac{1}{(2\pi M_{\textrm{cl}})^{(D-1)/2}}e^{i {\bf
      P}\cdot{\bf X}}\tilde \Psi_{\bP}(U^m)\;. 
\end{equation}
The $\tilde \Psi$ are wavefunctions on the centered moduli
space. In general we will denote quantities associated with the
centered part of the moduli space with a tilde.

 Translational invariance implies that the soliton solution
 depends on the center-of-mass moduli only through the difference
 ${\bf x - X}$, so that $\phi(\bx;U^M) = \phi(\bx - \bX;U^m)$. Using
 this fact, along with the factorized form of the wavefunctions, we
 have
\begin{align}\label{formfactor}
& \langle {\bf P}_f| e^{i H (t-T)} \phi(\bx;U) e^{-i H (t + T)}| {\bf P}_i\rangle = \cr
& \qquad =\frac{e^{-i (E_i + E_f) T} e^{-i (E_i - E_f )t}}{(2 \pi)^{D-1}} \int \ed \bX e^{i (\bP_i - \bP_f) \cdot \bX} \int_{\tilde{\MM}} \ed U \sqrt{\tilde{G}} \tilde{\Psi}_f^* \phi(\bx - \bX;U^m) \tilde{\Psi}_i (1+ O(g)) \cr
&  \qquad = \frac{e^{-i(E_i + E_f) T} e^{-i (P_i-P_f)\cdot x} }{(2\pi)^{D-1}}\int_{\tilde{\MM}}\ed U \sqrt{\tilde{G}} \tilde{\Psi}_f^* \FF[\phi] \tilde{\Psi}_i (1+ O(g))~,
\end{align}
where by $\mathcal F[ \phi ] =\mathcal F[\phi ]({\bP_i -
  \bP_f}; U^m) $ we denote the Fourier transform of $\phi$ with respect
to its first, spatial argument, and we have introduced the Lorentz spacetime momenta $P_{i,f} = (E_{i,f},\bP_{i,f})$.

Inserting \eqref{formfactor} into \eqref{leadingpert} and carrying out the integral over spacetime produces the energy-momentum conserving delta function, $(2\pi)^D \delta^{(D)}(k + P_i - P_f)$.  Hence, from \eqref{amplitude}, the leading semiclassical expression for the amplitude of
interest is given by a quantum mechanical matrix element of the
Fourier transform $\mathcal{F}[\phi]$ on the centered moduli space
\begin{align}
  \label{eq:1}
 {\cal A}(P_i, k\to P_f) = - i \frac{1}{(2 \pi)^{D-1}}e^{-i (E_i + E_f) T}  \int_{\tilde{\mathcal M}} \ed U \sqrt{\tilde
    G}\;\tilde\Psi_f^*\mathcal F[\mathcal \phi](-\bk;U^m) \tilde\Psi_i (1+ O (g))\;.
\end{align}
In the special case where there are no centered moduli,
\eqref{eq:1} reduces to the known result that the form factor
is just $\mathcal{F}[\phi]$, up to an energy-momentum-preserving
$\delta$-function \cite{Goldstone:1974gf} with standard initial and final state normalization factors.

Let us now comment on the validity of our calculation. Note that the result \eqref{eq:1} was obtained in the small-velocity approximation \eqref{manton}; the form of the energy eigenvalues \eqref{nrEvalue} provides a clear manifestation of the nonrelativistic limit. However, the regime needed to extract information about the pair-creation process through crossing symmetry requires large velocity exchange and hence momentum transfer of the order of the soliton mass.

In the case of the two-dimensional kink in $\Phi^4$ theory, seminal
work by Gervais, Jevicki and Sakita \cite{Gervais:1975pa} showed how
velocity corrections can be systematically accounted for to recover
the covariant expression for the soliton energy, $M_{\rm cl}\to \sqrt{
  {\bf P}^2 + M_{\rm cl}^2}$. This answer is to be expected, since the
starting point is a Lorentz-invariant theory. In the next section we
will show how the same techniques can be applied in the more general
class of Lorentz-invariant theories considered here. We will be
interested in evaluating the form factor \eqref{amplitude} rather than
the soliton energy.  In the next section we adapt the techniques of 
\cite{Gervais:1975pa,Dorey:1993by} to obtain an expression for the form factor that is
relativistic in the external momenta, $P_i, P_f$, but still only leading order in the ratio of the momentum transfer to the soliton mass.

\section{Evaluation of the relativistic form factor}\label{relativistic}

We now proceed to evaluate the form factor for processes involving
large velocity, \ie $P_{i,f}\sim O(M_{\rm cl})$. The path integral formulation is much more
appropriate for the purposes of resumming the relativistic corrections
and we will favor it over of the canonical approach implemented thus
far. The two qualitative differences between the general case and the
kink in $\Phi^4$ theory as considered by \cite{Gervais:1975pa} are,
first, a lack of an explicit classical soliton solution to work with
and, second, the possible presence of centered moduli. Both can be
taken into account and their discussion can be appropriately modified,
provided we continue to make the simplifying assumptions of the Manton
(small-velocity and small moduli-space-potential) approximation for
the dynamics of the centered moduli. Specifically, we will impose
\begin{align}
  \label{eq:21}
& P_m/m \sim O(1/g)~, \quad s(\bx;U^m) \sim O(1)\quad\textrm{and}\quad
\delta{\rm v}(U^m) \sim O(1)\cr
 & \textrm{but we take}\quad \bP/m \sim O(1/g^2)
\quad\textrm{such that}\quad \bP \propto M_{\rm cl}~.
\end{align}
This corresponds to a low energy restriction on the \emph{internal} states of the soliton that we consider.

The transition amplitude from an initial state $i$ described by $\Psi_i(U^M(-T);\chi)$ to a final state $f$ described by $\Psi_f(U^M(T);\chi)$ is
\begin{align}\label{transamp}
S_{fi} & =~ \int [DU DP D\chi D\pi] \delta(F_1) \delta(F_2) e^{i \int_{-T}^T \ed t' L} \Psi_{f}^\ast \Psi_i~, \quad \textrm{with} \cr
L & =~P_M \dot{U}^M + \int \ed \bx \;\pi \dot{\chi} - H~.
\end{align}
An incoming soliton state of momentum $\bP_i$ is defined, at leading order in the semiclassical expansion, by taking
$\Psi_i = \frac{1}{(2\pi)^{(D-1)/2}} e^{i \bP_i \cdot \bX_i}\tilde \Psi_{i}(U^m)$, where $\bX_i =
\bX(-T)$, and similarly for outgoing soliton states. We remind that the $\tilde
\Psi_{i,f}$ are normalized wavefunctions on the centered moduli space. We can consider time-ordered correlators of
the meson field between soliton states by inserting appropriate
factors of $\Phi(x_1)\cdots \Phi(x_n)$ under the path integral, and
using the relation $\Phi(x) = \phi(\bx - \bX(t);U^m) + g \chi(t,\bx -
\bX(t);U^m)$.

We are interested in the particular case of the 1-point function and
hence in 
\begin{align}\label{full1pt}
\langle \bP_f,T | \Phi(x) | \bP_i,-T \rangle =&~ \frac{1}{(2 \pi)^{D-1}}\int [D\bX D \bP] e^{i
  (\bX_i\cdot \bP_i - \bX_f\cdot \bP_f)} \int[DU^m DP_n] \tilde \Psi^*_f\tilde \Psi_i \times \cr
&~ \times \int [D \chi D \pi] \delta(F_1) \delta(F_2) e^{i \int_{-T}^T \ed t' L} \Phi[U,P;\chi](x)~.
\end{align}
Let us focus first on the internal path integral over $\chi$ and $\pi$
for which we will proceed to compute the leading contribution at small
$g$. This was done in \cite{Gervais:1975pa} for the case of the
transition amplitude by evaluating the action on the saddle-point solution
for $\chi,\pi$ corresponding to the moving soliton. We argue in
Appendix~\ref{app:sources} that the same saddle point solution gives
the leading contribution to the one-point function, even though one
should now be solving the equations of motion with source. This is a
special feature of working with the one-point function and would not
be true for higher point functions.  A completely analogous discussion
can be found in the papers of Dorey \emph{et.al,}
\cite{Dorey:1993by,Dorey:1994fk}. We denote this saddle point
$(\chi_{\rm cl}, \pi_{\rm cl})$ and expand the fields as $\chi =
\chi_{\rm cl} + \delta\chi$, $\pi = \pi_{\rm cl} + \delta \pi$.

Before continuing with the details, we wish to emphasize one property
of the semiclassical limit in which we work. One approach to the
computation would be the following. One could work around a saddle
point of the Hamiltonian \eqref{eq:3}, taking $g$ to be small, while
holding $\bP$ fixed---the soliton momenta enter \eqref{eq:3} through
\eqref{theas}. Thus, one would effectively be sending $g \to 0$,
carrying out the computation for arbitrary $\bP_{i,f}$, and then at
the end one could consider the limit of the result as the transfer
$\Delta \bP/m = (\bP_f - \bP_i)/m \to \infty$. (Here we inserted a
factor of the meson mass to get a dimensionless quantity.) However
this is not the correct limit to consider if one wishes to access the
theory in the regime related to pair creation by crossing symmetry.
Rather one should be sending $g \to 0$ and $\Delta \bP /m \to \infty$
{\it simultaneously}, while holding the velocity transfer $\Delta v
\sim \Delta \bP /M \sim g^2 \Delta \bP/m$ fixed and
$O(1)$.  Thus, it is important that the momentum is
treated as an $O(g^{-2})$ quantity---\ie\ the same order as the
soliton mass.  This is indeed what we do, and it is also an important,
but unstated, assumption in the original analysis of
\cite{Gervais:1975pa}. It is furthermore necessary to treat changes in the momentum---\ie time derivatives of $P$---as $O(g^{-2})$. This we do not do, which is why the following only holds at leading order in momentum transfer; see footnote~\ref{footnote}.\footnote{We thank E.~Witten for communication on this
  point.}

Now let us return to the
computation. Starting with the Hamiltonian \eqref{eq:3} one can find a saddle-point
solution to the $\chi,\pi$ equations of motion perturbatively in $g$
by making use of the scaling assumptions for the momenta \eqref{eq:21}. The details of this calculation are
carried out in Appendix~\ref{app:chipi}. One finds
\begin{equation}\label{chicl}
\chi_{\rm cl} = g^{-1} \phi\left( \Lambda(\bP) (\bx -
  \bX); U^m \right) - g^{-1} \phi\left((\bx -
  \bX); U^m \right) + O(1)~,
\end{equation}
where 
\begin{equation}\label{chisol3}
{(\Lambda[\bP])^i}_j = {\delta^i}_j + \left( \sqrt{1 + \frac{\bP^2}{M_{\rm cl}^2}} - 1 \right) \frac{P^i P_j}{\bP^2} ~
\end{equation}
are the spatial components of the Lorentz boost transformation corresponding to a relativistic momentum $\bP$. The insertion can then be expressed
as
\begin{align}
  \label{eq:9}
  \Phi  =   \phi\left(R^{-1}_S \Lambda[\bP] (\bx - \bX); U^m \right)+ O(g)
  \equiv \Phi_{\rm cl}(\bx-\bX,\bP;U_m) + O(g)\;.
\end{align}
The quantity $R_S$, inserted in the argument of the classical
soliton solution on dimensional grounds,
characterizes the size of the soliton. For example, in $\Phi^4$ theory
$R_S = 1/m$, with $m$ the meson mass.  As we previously indicated, in
the general class of theories considered here it can in principle be a
function of the centered moduli.

With this solution in hand, we want to evaluate \eqref{full1pt} in the
presence of centered moduli. For this, we also need the Lagrangian
evaluated on the solution:
\begin{equation}\label{fullL}
L = \bP \cdot \dot{\bX} - \sqrt{\bP^2 + M_{\rm cl}^2} + L^{(0)}[U^m, P_m; \delta\chi,\delta\pi; \bP] + L_{\rm int}~,
\end{equation}
where $L_{\rm int}$ starts at $O(g)$ and
\begin{equation}\label{laglag}
L^{(0)} = P_m \dot{U}^m - \tilde H_{\rm eff}[U^m,P_m;\bP]
\end{equation}
is an $O(1)$ contribution describing the dynamics of the centered
moduli, whose precise form we will not require. $\tilde H_{\rm eff}$
includes the 1-loop potential from integrating out the fluctuation
fields $(\delta\chi,\delta\pi)$. The leading contribution to
\eqref{full1pt} then takes the form
\begin{align}\label{foureight}
  & \langle \bP_f | \Phi(x) | \bP_i \rangle = \frac{1}{(2 \pi)^{D-1}} \int [D \bX D\bP] e^{i
    (\bX_i \cdot \bP_i - \bX_f \cdot\bP_f)} e^{i \int_{-T}^T \ed t'
    (\bP \cdot \dot{\bX} -
    \sqrt{ \bP^2 + M_{\rm cl}^2} ) } \times \cr &\times \int_{\tilde
    \cM} \ed U\sqrt{\tilde G}\; \tilde\Psi^*_f(U^m;\bP (T)) \Phi_{\rm
    cl}[\bx-\bX (t),\bP (t);U^m]\tilde\Psi_i(U^m;\bP (-T)) \left(1 + O(g)\right) .\quad
\end{align}
In the above we have expressed the centered moduli space path integral
as a position-basis matrix element in the quantum mechanics on the
centered moduli space with Hamiltonian $\tilde H_{\rm eff}$. 

The $(\bX,\bP)$ path integral is a functional integral representation
of the quantum mechanics for a relativistic particle. From the point
of view of the translational moduli space dynamics, $U^m$ are merely
parameters, so we can carry out the functional integration over $\bX$
and $\bP$ first and then integrate over the centered moduli space.
Thus the quantity that we would like to study is 
\begin{align}
  \label{eq:17}
  &\frac{1}{(2 \pi)^{D-1}}  \int [D \bX D\bP] e^{i (\bX_i\cdot \bP_i - \bX_f \cdot\bP_f)} e^{i \int_{-T}^T \ed t'
    (\bP \cdot \dot{\bX} - \sqrt{ \bP^2 + M_{\rm cl}^2} ) } \OO
  ~,\cr
& \textrm{where} \qquad \OO = \tilde\Psi^*_f(U^m;\bP (T)) \Phi_{\rm
  cl}[\bx-\bX (t),\bP (t);U^m]\tilde\Psi_i(U^m;\bP (-T))\;.
\end{align}
Making use of the Weyl--Wigner transform, the path integral can be carried out explicitly,  resulting in
\begin{align}
  \label{eq:19}
  \langle \bP_f | \Phi(x)|\bP_i\rangle =&~ e^{i (P_f - P_i)\cdot x}\int_{\tilde \cM} \ed U\sqrt{\tilde
    G}\;\tilde\Psi_f^* \FF[ \phi] \left( R_S\;{\left(\Lambda^{-1}\left[\tfrac{\bP_i+\bP_f}{2}\right]\right)^i}_j(P_i - P_f)^j\right) \tilde\Psi_i \times \cr
    &~ \times \frac{e^{-i (E_i + E_f)T}R_{S}^{D-1}}{(2\pi)^{D-1}\sqrt{1 + (\tfrac{\bP_i + \bP_f}{2 M_{\rm cl}})^2} } \, (1 + O(g))~,  \raisetag{24pt}
\end{align}
where $\FF[\phi](u) =\int \ed v\;  e^{- i u v} \phi(v)$ is the Fourier transform of the classical
soliton profile, $  \tilde\Psi_{i,f} = \tilde\Psi_{i,f}
(U^m;P_{i,f})$, and $\Lambda^{-1}$ is the inverse of the matrix given in \eqref{chisol3}.

Hence, taking into account the parameterization \eqref{chisol3} of the classical soliton profile, the generalization of the amplitude, \eqref{eq:1}, to relativistic initial and final momenta is
\begin{align}\label{RelAmp}
\mathcal{A}(P_i,k \to P_f) =&  -i \int_{\tilde \cM} \ed U\sqrt{\tilde
    G}\;\tilde\Psi_f^* \FF[ \phi] \left( R_S\;{\left(\Lambda^{-1}\left[\tfrac{\bP_i+\bP_f}{2}\right]\right)^i}_j(P_i - P_f)^j\right) \tilde\Psi_i  \cr
    &~ \times \frac{e^{-i (E_i + E_f)T}R_{S}^{D-1}}{(2\pi)^{D-1}\sqrt{1 + (\tfrac{\bP_i + \bP_f}{2 M_{\rm cl}})^2} } \, (1 + O(g))~.  \raisetag{24pt}
\end{align}
This result treats the initial and final soliton momenta relativistically, but is based on a saddle point solution of the soliton sector equations of motion with constant soliton momentum $P$.  Therefore \eqref{RelAmp} only captures the leading order behavior of the semiclassical form factor in the ratio of the momentum transfer to the soliton mass, $k/M_{\rm cl}$.

\section{Comments on the Analytic Continuation to the Pair Production Amplitude}

Equation \eqref{RelAmp} cannot be used to analytically continue to the production amplitude, as it is not valid in the kinematic region, $k \sim O(M_{\rm cl})$, required for pair production.  A related observation is that the expression is not Lorentz-invariant.  This is to be expected given that it is only a leading order expression in the momentum transfer.

The proper way to proceed is via the analysis of \cite{Melnikov:2020ret}, which gives a solution in principle to all orders in momentum transfer.  However explicit computations require understanding time-\emph{dependent} solutions to the saddle-point equations discussed in Appendix \ref{app:chipi}.  The resulting semiclassical form factor is expected to be Lorentz invariant, as the original quantum field theory is Lorentz invariant and this property must hold order by order in the coupling expansion.

Having said that, we have found that various sources invoke the Breit (or brick wall) frame to provide a Lorentz covariantization of the nonrelativistic (or small-momentum-transfer) form factor and claim its validity for any value of the momentum transfer. Let us expand upon this point and various implementations of the Breit frame. The latter, defined by $\bP_f = - \bP_i = \bk/2$, is the unique frame for which the magnitude of the spatial momentum is the same as the magnitude of the Lorentz momentum, $|\bk | = \sqrt{- k^2}$.
\begin{itemize}
\item [i)]
  One may be tempted to express the {\it nonrelativistic} result \eqref{eq:1} in terms of the Lorentz-invariant quantity
\begin{align}
  \label{eq:5}
k^2 = (P_f-P_i)^\mu(P_f-P_i)_\mu \;,
\end{align}
which in the small velocity approximation leads to 
\begin{align}
  \label{eq:6}
 \sqrt{-k^2} = |\bk| \Big( 1 + O( \bP_i^2/M_{\rm cl}^2, {\bP}_f^2/M_{\rm cl}^2) \Big) \;,
\end{align}
and assume that making the replacement $|\bk| \to \sqrt{-k^2}$ correctly captures all relativistic corrections to \eqref{eq:1}.  If this were true, then one could extend the result \eqref{eq:1} to large spacelike $k^2$, corresponding to large values of the argument of the Fourier transform.  Assuming smoothness of the classical soliton profile, the Riemann--Lebesgue lemma would imply that the Fourier transform falls off faster than any power.  Then one could consider the analytic continuation from spacelike to timelike $k^2$ and attempt to draw a conclusion about the pair production process, as in \cite{Banks:2012dp}.  However, this leads to form factors that fall off faster than any power in momentum transfer. This is a result that is qualitatively incompatible with expectations from any asymptotically free theory, where the large momentum behavior of amplitudes is expected to be power-law.  See \eg \cite{Ji:1991ff} for a discussion in the context of skyrmion form factors in QCD.
\item[ii)]
The relativistic result, \eqref{eq:19}, can be evaluated in the Breit frame, where the Lorentz factor reduces to an identity matrix due to $\bP_i+\bP_f = 0$. Hence this gives an answer that is equivalent to the ``replacement rule,'' $|{\bf k}| \to \sqrt{-k^2}$ applied to the nonrelativistic form factor. This not only has the wrong asymptotic behavior in the momentum transfer, but when specializing to the case of sine--Gordon theory for $D=2$---where exact results can be obtained using integrability---it can be explicitly seen that it leads to the wrong answer \cite{Weisz:1977ii}.
\item[iii)]
Another approach that can be found in the literature is to boost the nonrelativistic form factor to the Breit frame.  More precisely, one boosts the static soliton profile to the Breit frame using a Lorentz transformation associated with the relativistic momentum $\bk/2$.  Next, one takes the form factor to be the Fourier transform of this boosted profile with respect to its position argument, using dual momentum $\bk$.  Finally, one sets $|\bk| = \sqrt{-k^2}$.  Provided that the original soliton profile only depends on the magnitude of its spatial argument, $|\bx - \bX|$, then the argument of the Fourier transform will involve
\begin{equation}
R_S \left| \left(\Lambda^{-1}[\tfrac{\bk}{2}]\right)^{i}_{~j} k^j \right| = \frac{R_S |\bk|}{\sqrt{1 + (\tfrac{\bk}{2M_{\rm cl}})^2}} = \frac{2 R_S}{R_C} \sqrt{\frac{ k^2}{k^2 - 4 M_{\rm cl}^2}} ~,
\end{equation}
where we have used $R_C = 1/M_{\rm cl}$.  Although this procedure leads to a Lorentz-invariant completion of \eqref{eq:1} with improved asymptotic behavior in the momentum transfer, no justification is given for choosing the Breit frame---or any frame for that matter. For an application in the context of the Skyrme model see \eg \cite{Ji:1991ff,Holzwarth:1996xq}. 
\item[iv)]
Finally, we note a curiosity related to the previous point.  In $D = 2$ spacetime dimensions, one could view the insertion 
\begin{align}
\Phi_{\rm cl} =&~  \phi\left(R_{S}^{-1} \Lambda[\bP] (\bx - \bX)\right) = \phi\left(R_S \sqrt{1 + \tfrac{P^2}{M_{\rm cl}^2}} \, (x - X) \right)~,
\end{align}
that appears in \eqref{eq:17}, as a gauge-fixed form of the insertion
\begin{equation}
\phi\left(\tfrac{R_C}{R_S} \epsilon_{\mu\nu} P^\mu (x-X)^\nu \right) \quad  \xrightarrow{\textrm{static gauge}} \quad \phi\left(\tfrac{R_C}{R_S} P^0 (x - X) \right)\;,
\end{equation}
in a manifestly Lorentz-invariant approach to the path integral of the relativistic particle.  In such an approach, one allows the time parameter, $X^0$, along the worldline of the particle to be independent of the coordinate time $t$, and mods out by reparameterizations of the worldline.  Then \eqref{eq:17} would correspond to choosing the static gauge, $t = X^0$.  Utilizing techniques in \cite{Mannheim:1985gs,Monaghan:1986by} to carry out the path integral from the Lorentz-invariant starting point, one finds that the form factor is given by the Fourier transform of the classical profile, with argument $\frac{2 R_S}{R_C} \zeta(P_f, P_i)$, where 
\begin{align}
  \zeta(P_f, P_i) := \frac{2 \epsilon_{\mu\nu}P_f^\mu P^\nu_i}{(P_f + P_i)^2}\, .
\end{align}
Using the mass shell conditions for $P_{i,f}$, one can check that
\begin{equation}
\zeta = \sqrt{\frac{ k^2}{k^2 - 4 M_{\rm cl}^2}} ~,
\end{equation}
the same quantity that appears when boosting the static profile to the Breit frame, as discussed under item iii).

This expression agrees with \eqref{RelAmp} in the low momentum transfer limit, but the two disagree away from this limit.  This is not unexpected, since the manifestly Lorentz invariant formulation of the relativistic particle path integral and the form appearing in \eqref{eq:17} are known to give different results in the presence of insertions
\cite{Teitelboim:1980my,Henneaux:1982ma}.  Furthermore, ordinary relativistic quantum mechanics suffers well-known problems in the presence of interactions---problems that are resolved by quantum field theory \cite{Chen:2018cts}---so one should certainly not expect the covariantization procedure just discussed to reproduce the correct form factor. And indeed, one can again check against known results for the sine-Gordon model that it does not.  

\end{itemize}

In the absence of an explicit expression for the semiclassical form factor valid to all orders in momentum transfer, and despite the above comments, we find it is useful to take the expression resulting from points iii) or iv) as a toy model for investigating the analytic continuation to the pair production amplitude.  Therefore, ignoring $g$-independent normalization factors such as those appearing in \eqref{RelAmp}, and focusing on $D = 2$, we consider the amplitude
\begin{align}
  \label{eq:10}
  \cA_{\rm toy}(P_i, k \to P_f) \propto &~ \int_{\tilde \cM} \ed U\sqrt{\tilde
    G}\;\tilde\Psi_f^* \FF[ \phi] \left(
    \frac{2R_S(U^m)}{R_C}\zeta(P_f, P_i) \right) \tilde\Psi_i\;.
\end{align}
Here we have emphasized that the size parameter, $R_S$, will in general be a function of the centered moduli.  

Given that the classical soliton profile $\phi$ is a smooth $(C^\infty)$ function of its position argument, we can draw a rather strong conclusion about the asymptotic behavior of the Fourier transform in \eqref{eq:10}. For any values of momenta such that $\zeta$ is not $O(g^2)$ or smaller, it is the $2R_S/R_C$ factor that controls the parametric size of the argument of the Fourier transform.  Given this, and as long as the soliton size is bounded away from zero, $R_S^{\rm min}>0$, we will have that $(2R_S/R_C)|\zeta|\to\infty$ in the semiclassical limit. Here we remind the reader that $R_C^{-1} = M_{\rm cl}$ is assumed to be $O(1/g^2)$ relative to some fixed mass scale determined from the potential (\eg the meson mass). The Riemann--Lebesgue lemma then implies that\footnote{As stated by the
  Riemann-Lebesgue lemma, the Fourier
  transform $\mathcal F[f](p)$ of an $L^1$-function $f(x)$ goes to
  zero as $|p| \to \infty$. Accordingly, if $f(x)$ is $C^\infty$,
  $\mathcal F[f^{(n)}](p) = (ip)^n \mathcal F[f](p)$ should also go to
  zero as $p\to \infty$; \ie $\mathcal F[f](p)$ goes to zero faster
  than any power.}
\begin{equation}\label{Riemann}
\mathcal F[\phi ]\left(\frac{2R_S(U^m)}{R_C} \zeta\right)\sim  \;e^{-
  \frac{2R_S(U^m)}{R_C} |\zeta|} \quad \textrm{as} \quad
(2R_S/R_C)|\zeta|\to\infty \;.
\end{equation}
Let us emphasize that the exponential on the right-hand side is a typical function
exhibiting a faster-than-any-power falloff. It is used for
concreteness, but the exact expression will depend on the details of
the theory under consideration. In any case, the important property
for our purposes is the faster-than-any-power falloff in the coupling. 

This leads to the asymptotic estimate
\begin{align}
  \label{eq:14}
  \cA_{\rm toy}(P_i,k \to P_f) \sim& \int_{\tilde \cM} \ed U\sqrt{\tilde
    G}\;\tilde\Psi_f^* e^{- \frac{2R_S(U^m)}{R_C} |\zeta|} \tilde\Psi_i
\end{align}
for the leading contribution to the toy-model form factor as $g\to 0$. Note that
the centered moduli space represents the internal degrees of freedom
of the single-particle state. A field theory interpretation requires a
single-particle state to have a finite number of internal degrees of
freedom. The eigenvalues labeling them should be discrete eigenvalues
of the centered-moduli-space Hamiltonian $\tilde H_{\rm eff}$.  Hence
the wavefunctions on the centered moduli space $\tilde{\Psi}$ should
be $L^2$; this is automatically the case if $\tilde{\mathcal M}$ is
compact. Then we have the inequalities
  \begin{align}
    \int_{\tilde \cM} \ed U\sqrt{\tilde G}\;\tilde\Psi_f^* e^{-
      \frac{2R_S(U^m)}{R_C} |\zeta|} \tilde\Psi_i & \le \int_{\tilde
      \cM} \ed U\sqrt{\tilde G}\;|\tilde\Psi_f^* \tilde\Psi_i | e^{-
      \frac{2R_S(U^m)}{R_C} |\zeta|} \cr & \le e^{- \frac{2R_S^{\rm
          min}}{R_C} |\zeta|} ||\tilde\Psi_f^* \tilde\Psi_i
    ||_{L^1}\cr & \le e^{- \frac{2R_S^{\rm min}}{R_C} |\zeta|}
    ||\tilde\Psi_f||_{L^2} ||\tilde\Psi_i ||_{L^2}\cr & = e^{-
      \frac{2R_S^{\rm min}}{R_C} |\zeta|} \;,
  \end{align}
where in the second-last step we used H\"older's inequality. Hence we have reached the result
\begin{align}\label{finalA}
  \cA_{\rm toy}(P_i,k \to P_f) \lesssim e^{- \frac{2R_S^{\rm min}}{R_C} |\zeta|}\;. 
\end{align}

Although the toy-model form factor is not the same as the actual field-theoretic one, the above calculation shows that the large-$k$ behavior of the right-hand side of \eqref{eq:10} is consistent with expectations for the true form factor. As $k^2\to \infty$ we expect $\zeta(k^2)\to O(1)$; otherwise, one would obtain an amplitude with exponential behavior for large $k^2$, in contradiction with the large-momentum behavior of asymptotically free theories.

If one did have the correct semiclassical form factor to all orders in $k/M_{\rm cl}$, at this stage one would analytically continue from spacelike to timelike $k^2$ to obtain the pair production amplitude. It is interesting to observe that if one were to perform this operation directly on \eqref{finalA}, one would conclude that the pair production amplitude is exponentially suppressed in $2R_S^{\rm min}/ R_C$ at large $k$. Note that if $R_S^{\rm min}$ is
of order $R_C$ this does not lead to suppression.

\section{Instanton-solitons in 5D MSYM}\label{instantons}

Extending the discussion to more
general Lagrangians with gauge fields and fermions introduces
technical complications related to gauge invariance (gauge zero modes,
ghosts). However, the simplicity of the idea compels us to apply this framework to the interesting case of
instanton-solitons in maximally supersymmetric 5D Yang--Mills (MSYM)
theory.

Yang-Mills theory in 5D is normally viewed as an effective field
theory, valid at low energies. However, the connection of 5D MSYM to
the $(2,0)$ SCFT in 6D leaves open the possibility that this theory is
in fact well defined \cite{Douglas:2010iu,Lambert:2010iw}, even though
it is perturbatively divergent at six loops
\cite{Bern:2012di}.\footnote{Some nontrivial results compatible with
  this conjecture include \cite{Tachikawa:2011ch, Lambert:2011eg,
    Kim:2011mv, Young:2011aa, Kallen:2012cs, Hosomichi:2012ek,
    Kallen:2012va, Kim:2012av, Kallen:2012zn, Bak:2012ct,
    Fukuda:2012jr, Kim:2012qf, Lambert:2012qy, Bak:2013bba}.} In line
with the rest of this letter, we argue that {\it before} sending the
cutoff to infinity and declaring 5D MSYM to be UV-divergent, one has
to also investigate contributions associated with soliton-antisoliton
pair production. We stress that in doing so we are not treating this
theory as an effective theory in the Wilsonian sense.

Instanton-solitons in 5D MSYM are finite-energy $\half$-BPS field
configurations. They solve the selfduality equation for the gauge
field strength in the four spatial directions and as such are
described by conventional 4D instanton solutions. For topological
charge $c_2(F) =1$ and $\SU(2)$ gauge group the classical gauge field
is given by
\begin{equation}
  A_i = U(\vec
  \theta)^{-1}\Big(\frac{\eta_{ij}^a(\bx-\bX)^j}{(\bx-\bX)^2 +
    \rho^2}T^a\Big)U(\vec\theta)~, \qquad A_0 = 0\;,
\end{equation}
with $a = 1,2,3$, $i = 1,...,4$ and $\eta_{ij}^a$ the `t Hooft
symbols. This solution has eight moduli: four center-of-mass
collective coordinates $\bX$, a size modulus $\rho$ and three Euler
angles $\vec \theta$ parameterizing global gauge transformations.  The
associated moduli space is a hyperk\"ahler manifold
\begin{equation}
 \mathcal M = \mathbb R^4 \times \mathbb R_+ \times S^3/\mathbb Z_2\;,
\end{equation}
with metric
\begin{equation}
 \ed s^2 = \frac{4\pi^2}{g_{\rm YM}^2}\Big[\delta_{ij}\ed \bX^i \ed\bX^j +
 2(\ed\rho^2+ \rho^2 \tilde G_{\alpha\beta}\ed\theta^\alpha \ed\theta^\beta)\Big] \;,
\end{equation}
where $\tilde G_{\alpha\beta}$ is the metric on ${\rm SO}(3) \cong
S^3/\mathbb Z_2$, the group of effective global gauge transformations,
and $g^2_{\rm YM}$ has dimensions of length. 

The presence of a noncompact size modulus translates into
instanton-solitons that can have arbitrarily small or large
sizes. A naive application of the ideas discussed in the introduction would then imply that
instanton-soliton pair production need not be suppressed relative to
perturbative processes. However, this non-compact direction in the
centered moduli space $\tilde\cM $ also results in the corresponding
Hamiltonian not admitting $L^2$-normalizable eigenfunctions. The
centered Hamiltonian does possess a continuum of plane-wave
normalizable wavefunctions, but this renders the interpretation of
instanton-solitons as asymptotic states confusing, since they would
correspond to particles with a continuously infinite number of
internal degrees of freedom.

Moreover, the parameter controlling the semiclassical expansion of the
Hamiltonian in the soliton sector is in fact $g^2 = g_{\rm
  YM}^2/\rho$, which coincides with $R_C/R_S$. In particular, note
that $g=g(\rho)$ is now moduli dependent. In the context of the
semiclassical expansion \eqref{Hamexp}, or more appropriately the
relativistic version in Appendix~\ref{app:chipi}, we can imagine a
fixed $\rho$ such that $g(\rho)$ is small. However, when evaluating
amplitudes, where one must integrate over all sizes, the semiclassical
approximation breaks down. Consequently, the small-sized
instanton-solitons invalidate our argument for exponential
suppression.

One can attempt to circumvent this conclusion by turning on a scalar
VEV, $\langle \Phi \rangle \neq 0$, and going out onto the Coulomb
branch.\footnote{Here $\Phi$ is one of the five adjoint scalars of 5D
  SYM and should not be confused with the scalar fields for the linear
  sigma models considered in the previous sections.}  It is known that
in this case finding instanton-soliton solutions requires turning on
an electric field, which stabilizes the \emph{classical} size
\cite{Lambert:1999ua,Peeters:2001np,Allen:2012rp}. From the point of
view of the quantum theory, turning on an electric field generates a
potential on the centered moduli space,
\begin{equation}\label{potential}
\delta {\rm v}(U^m) = \frac{2 \pi^2}{g_{\rm YM}^2} \langle \Phi \rangle^2 \rho^2~,
\end{equation}
and lifts the flat direction associated with the instanton-soliton
size.  Although $\rho$ is no longer a true modulus, the VEV provides
an additional dimensionless parameter, $ \epsilon := g_{\rm YM}^2
\langle\Phi \rangle$, that can be adjusted so that we remain in the
small-potential approximation \eqref{eq:21}, where it is still
appropriate to represent states as $L^2$-wavefunctions on
$\tilde{\mathcal{M}}$. In order to determine the precise form of the
resulting $L^2$-wavefunctions, one would need to compute the
centered-moduli-space Hamiltonian $\tilde H_{\rm eff}$, appearing in
\eqref{fullL} and \eqref{laglag}.\footnote{One will actually have a
  supersymmetric quantum mechanics with 8 supercharges so the
  wavefunctions will be forms or bispinors on the moduli space, due to
  realizing the fermi collective coordinate anticommutator as a
  Clifford algebra \cite{Bak:1999da}.}

Our formalism has been general enough to accommodate such potentials
on moduli space. Thus, despite the classical stabilization, one must
still integrate over all of moduli space, which includes arbitrarily
small sizes. However, as we have already discussed, this means
treating the solitons semiclassically when $\rho\sim O (g_{\rm
  YM}^2)$, which is not valid because quantum corrections that have
been neglected become important. Hence, turning on the potential
\eqref{potential} does not enable one to salvage an argument for
faster-than-any-power suppression.

While none of these arguments definitively show that
instanton-soliton contributions are {\it not} suppressed compared to
perturbative processes, they at least allow for that
possibility. Non-suppression of the pair-production amplitude would
provide a mechanism via which the contribution of virtual
soliton-antisoliton pairs to perturbative processes such as
\eqref{optical} can compete with the contribution from loops of
perturbative particles.  Such a mechanism is precisely what is called
for in order to avoid contradicting the assumption of finiteness: One
would require that the soliton-antisoliton contribution be divergent,
with exactly the right coefficient to cancel the divergence found in
\cite{Bern:2012di}.  This is an intriguing possibility, the
investigation of which would, however, require an alternative approach
to the one used here.

\newpage

\section*{Acknowledgments}

We would like to thank Philip Argyres, Tom Banks, Shabnam Beheshti,
Jacques Distler, Michael Douglas, Daniel Friedan, Jeff Harvey, Daniel
Jafferis, Seok Kim, Ilarion Melnikov, Greg Moore, John Schwarz and
Edward Witten helpful discussions and comments. CP is a Royal Society
Research Fellow and partly supported by the U.S. Department of Energy
under grants DOE-SC0010008, DOE-ARRA-SC0003883 and
DOE-DE-SC0007897. ABR is supported by the Mitchell Family
Foundation. We would like to thank the Mitchell Institute at Texas
A\&M and the NHETC at Rutgers University respectively for hospitality
during the course of this work.  We would also like to acknowledge the
Aspen Center for Physics and NSF grant 1066293 for a stimulating
research environment.

\appendix

\section{One-point function and the source-free e.o.m.}\label{app:sources}

Suppose we add sources $(J,K)$ for $(\chi,\pi)$ to the Lagrangian \eqref{transamp}:
\begin{equation}
L \to L + \epsilon \int \ed \bx \left( J \chi + K \pi \right)~.
\end{equation}
Here we have introduced a small parameter $\epsilon$; we will solve
the classical equations of motion with source perturbatively in
$\epsilon$. It should be sufficient to treat the sources in this
fashion since we are only interested in $n$-point correlators for
which one only needs the behavior of the partition function in a
neighborhood of $J= 0 = K$.  We make a series expansion
\begin{equation}
\chi_{\rm cl} = \chi_{\rm cl}^{(0)} + \epsilon \chi_{\rm cl}^{(1)} + O(\epsilon^2)~, \qquad \pi_{\rm cl} = \pi_{\rm cl}^{(0)} + \epsilon \pi_{\rm cl}^{(1)} + O(\epsilon^2)~,
\end{equation}
and plug into the classical equations of motion with source. We then
have to expand in fluctuations around the classical solution, $\chi =
\chi_{\rm cl} + \delta \chi$.  The contributions from the fluctuations
are suppressed in powers of $g$.

Demanding that the classical equations hold order by order in
$\epsilon$, we find that $(\chi_{\rm cl}^{(0)},\pi_{\rm cl}^{(0)})$
should solve the source-free equations of motion. The $(\chi_{\rm
  cl}^{(1)},\pi_{\rm cl}^{(1)})$ solve an inhomogeneous linear
differential equation, involving the operator that controls the
spectrum around the soliton. Let us denote the restriction of that
operator to the space orthogonal to the zero modes by $\OO$ (so that
$\OO$ has an inverse). We assume that $(J,K)$ have no overlap with the
zero modes. Then the solution to $O(\epsilon^2)$ is
\begin{equation}
\left(\begin{array}{c} \chi_{\rm cl} \\ \pi_{\rm cl} \end{array}
\right) = \left(\begin{array}{c} \chi_{\rm cl}^{(0)} \\ \pi_{\rm
      cl}^{(0)} \end{array} \right) + \epsilon \OO^{-1} \left(\begin{array}{c} J \\ K \end{array} \right) + O(\epsilon^2)~.
\end{equation}
Now we must plug this back into the Lagrangian with source.  Notice
that, crucially, because $(\chi_{\rm cl}^{(0)}, \pi_{\rm cl}^{(0)})$
satisfy the source-free equations of motion, the original Lagrangian
has an expansion $L[\chi_{\rm cl},\pi_{\rm cl}] = L_{\rm cl}^{(0)} +
O(\epsilon^2)$; there are no linear terms in the source.  Similarly,
since we are expanding around a solution $(\chi_{\rm cl},\pi_{\rm
  cl})$ to the equations of motion, the first corrections from the
quantum fluctuations come at quadratic order in $(\delta\chi,
\delta\pi)$.  Hence, the only linear terms in the source come from
$(\chi_{\rm cl}^{(0)} + \delta\chi, \pi_{\rm cl}^{(0)} + \delta\pi)$
multiplying $(J,K)$ in the source term itself:
\begin{align}
L + \int \ed \bx \left( J \chi + K \pi \right) =&~ L_{\rm cl}^{(0)} + \int \ed \bx \left( J (\chi_{\rm cl}^{(0)} + \delta\chi) + K (\pi_{\rm cl}^{(0)} + \delta\pi) \right) + \cr
&~ + O(J^2,K^2,\delta\chi^2, \delta\pi^2)~,
\end{align}
where we have reabsorbed $\epsilon$ into the sources.  

Hence, the one-point function for \eg the $\chi$-field is
\begin{align}
& \frac{\delta}{\delta J} \left( \int [D\chi D\pi] \delta(F_1)
  \delta(F_2) e^{i \int \ed t' (L + J\chi + K \pi)} \right) \bigg|_{J = 0 = K} = \cr
& \qquad \qquad =  \int [D\chi D\pi] \delta(F_1) \delta(F_2)
(\chi_{\rm cl}^{(0)} + \delta\chi) e^{i \int \ed t' ( L_{\rm cl}^{(0)} + O(\delta\chi^2, \delta \pi^2) ) } \cr
& \qquad \qquad = e^{i \int \ed t' L_{\rm cl}^{(0)}} \chi_{\rm cl}^{(0)}
\int [D \delta\chi D \delta\pi] \delta(F_1[\delta\chi])
\delta(F_2[\delta\pi]) e^{O(\delta\chi^2,\delta\pi^2)} \left( 1 +
  \delta\chi/\chi_{\rm cl}^{(0)} \right)~,\cr
\end{align}
where the ratio $\delta\chi/ \chi_{\rm cl}^{(0)}$ is $O(g)$. The
Gaussian integral over the fluctuations gives a one-loop correction to
$L_{\rm cl}^{(0)}$, which also only depends on $\chi_{\rm cl}^{(0)}, \pi_{\rm
  cl}^{(0)}$. As a result, the leading saddle-point contribution to the one-point function is expressed entirely in terms of the classical solution to the \emph{source-free} equations of motion.

\section{The $\chi,\pi$ saddle point in the presence of centered moduli}\label{app:chipi}

In this appendix we would like to find the saddle-point solutions to
the $\chi,\pi$ equations of motion coming from the
Hamiltonian\footnote{The saddle-point solutions that we will find here
  are the classical piece of the $\chi,\pi$ fields, previously denoted
  by $\chi_{\rm cl},\pi_{\rm cl}$. We will drop the subscripts in the
  following equations for brevity.}
\begin{align}\nn
H =&~   \frac{g^4}{2}  a^M
 G_{MN}  a^N  + {\rm v}(U)  +\int \Big[\half
   \pi \cdot \pi + g\; s\cdot  \chi + 
\frac{1}{2}  \chi \cdot \Delta  \chi + V_I(
\chi)  \Big]  + O(g^2) ~, 
\end{align}
with
\begin{align}\label{sourceOp:app}
  & s(\bx;U^m) :=\frac{1}{g^2} \Big( - \pd_{\bx}^2 \phi + \frac{\pd V}{\pd \Phi}
  \bigg|_{\Phi = \phi} \Big)~, \qquad
  \Delta  := -\delta_{ab} \pd_{\bx}^2 + \frac{
  \delta^2 V}{\delta \Phi \delta\Phi}\bigg|_{\Phi = \phi}~, \cr 
 & {\rm v}(U^m)  := \frac{1}{g^2} \int \ed\bx \left(
    \half \pd_{\bx} \phi 
    \cdot \pd_{\bx} \phi + V(\phi) \right) = M_{\rm cl} + \delta
  {\rm v}(U^m)\;.
\end{align}
The equations of motion are given by
\begin{equation}
\dot{\chi} = \frac{\delta H}{\delta \pi} - \nu^M \pd_M \phi~,
\qquad \dot\pi = - \frac{\delta H}{\delta \chi} - \mu^M \pd_M \phi~,
\end{equation}
where the $\mu,\nu$ are Lagrange multipliers for the second-class
constraints
\begin{align}
  \label{eq:16}
  F_{1,M} := \int \chi \cdot \pd_M \phi = 0\;, \qquad F_{2,M} := \int \pi \cdot \pd_M \phi = 0~.
\end{align}
The equations we want to solve are
\begin{align}
0 =&~ \dot{\chi} + \nu^M \pd_M \phi - \pi + \pd_M \chi (C G C)^{MN} (P_N - \smallint \pi \cdot \pd_N \chi)~, \cr
0 =&~ \dot\pi + \mu^M \pd_M \phi + \frac{1}{g} \pd_{\bx}^2 (\phi + g
\chi) -\frac{1}{g} V'(\phi+g\chi) + \cr
& ~+ g \;s - \pd_M \pi(C G C)^{MN} (P_N - \smallint \pi \cdot \pd_N \chi) + \cr
&~ -g^{-1} (\pd_M \pd_N \phi) C^{MP} (P_P - \smallint \pi \cdot \pd_P \chi) (C G C)^{NQ} (P_Q - \smallint \pi\cdot \pd_Q \chi)~,
\end{align}
where $ C^{MN} = [(G - g \Xi)^{-1}]^{MN}$ and
\begin{align}
  \label{eq:15}
    G_{MN} := \frac{1}{g^2}\int \pd_M \phi\cdot \pd_N \phi \;,\qquad
  \Xi_{MN} := \frac{1}{g^2}\int \chi\cdot \pd_M \pd_N \phi\;.
\end{align}

Our strategy will be to work in the following approximation scheme. On
the one hand, after changing variables to the soliton-fixed frame,
$\rho = \bx - \bX$, one finds that the equations do not depend on
$\bX$ and thus it is consistent to treat $\bP$ as constant.\footnote{This is true provided we do not consider $X$ dependent insertions in the transition amplitude. However, the form factors we are interested in do involve $X$ dependent insertions, and by taking $P$ constant here we can only recover the leading order behavior of these form factors in the momentum transfer; see footnote~\ref{footnote}.} On the
other, the equations do depend on the relative moduli $U^m$ (through
the dependence of the metric and the fields on them) so it is
inconsistent to treat $P_m$ as constant. However, we will assume that
the motion of the relative moduli is slowly varying, $\dot{U}^m \sim
O(g)$, so that altogether
\begin{align}\label{momscale}
& \bP \sim O(g^{-2})~, \quad \dot{\bP} = 0~, \quad P^m \sim O(g^{-1}) \cr & s(\bx;U^m) \sim O(1) \quad \Rightarrow
  \quad M_{\rm cl} \sim O(1/g^2) ~ \textrm{ and } ~ \delta {\rm v}(U^m) \sim
  O(1)~.
\end{align}
For example, note that in this approximation $\dot{\chi} = \pd_m \chi
\dot{U}^m$ is suppressed by a factor of $g$ relative to $\chi$, since
$\dot{U}^m \sim O(g)$.

We first solve for $\nu^M$:
\begin{align}
  \nu^M = G^{MN} (\Xi)_{NP} (C G C)^{PQ} (P_Q - \smallint \pi \cdot \pd_Q \chi) - G^{MN} \frac{1}{g^2} \int \pd_N \phi \cdot \dot{\chi}~,
\end{align}
and then use that to find $\pi$:
\begin{align}\label{pi0sol2}
\pi =&~ \Pi^\perp \dot{\chi} + \pd_M \phi  G^{MN} (\Xi)_{NP} (C G C)^{PQ} (P_Q - \smallint \pi \cdot \pd_Q \chi) + \cr
&~ + \pd_M \chi (C G C)^{MN} (P_N - \smallint \pi \cdot \pd_N \chi)~,
\end{align}
where
\begin{equation}
\Pi^\perp \dot{\chi} := \dot{\chi} - \pd_M \phi G^{MN} \frac{1}{g^2} \int \pd_N \phi \cdot \dot{\chi}
\end{equation}
is the projection of $\dot{\chi}$ to the space orthogonal to the
zero modes. Expressing $G^{-1} \Xi = g^{-1} (\mathbbm{1} - G^{-1} C)$,
we can rewrite the above equation as
\begin{align}
\pi =&~ \Pi^\perp \dot{\chi} + \pd_M (g^{-1} \phi + \chi) (C G C)^{MN} (P_N - \smallint \pi \cdot \pd_N \chi) + \cr
&~ - g^{-1} \pd_M \phi (G^{-1} C^{-1})^{M}_{\phantom{M}P} (C G C)^{PN} (P_N - \smallint \pi \cdot \pd_N \chi)~.
\end{align}
We will then assume our solution for $\chi$ is of the form
\begin{equation}\label{B11}
  \chi = \tilde{\chi}^{(-1)} - g^{-1} \phi + \tchi^{(0)}~,
\end{equation}
where $\tchi^{(-1)}$ and $\tchi^{(0)}$ are $O(g^{-1})$ and $O(1)$
terms respectively. Then the solution for $\pi$ can be expressed to
leading order as
\begin{align}\label{piinter}
  \pi =~ &  \Pi^\perp \dot{\tchi}^{(-1)} + \left( \pd_M
    \tchi^{(-1)} + \pd_M \tchi^{(0)} - g^{-1} \pd_P \phi (G^{-1}
    C^{-1})^{P}_{\phantom{P}M} \right) D^{MN}\times \cr
&\times \left( P_N - \smallint \Pi^\perp \dot{\tchi}^{(-1)} \cdot \pd_N \tchi^{(-1)} \right) + O(g)~. 
\end{align}
Here we have defined a `relativistic' moduli space metric
\begin{equation}
  D_{MN} = \int \pd_M (\tchi^{(-1)} + \tchi^{(0)}) \cdot \pd_N (\tchi^{(-1)} + \tchi^{(0)})
\end{equation}
and in \eqref{piinter} it is understood that we only keep $D^{MN}$,
$C^{MN}$ to the appropriate order in $g$, denoted \eg by $D^{(n)MN}$.

Now let us turn to the $\chi$ equation, wich after various
manipulations can be written as
\begin{align}\label{Echihybrid}
E_{\chi} :=&~ \left[ \pd_t (\Pi^\perp \pd_i \tchi^{(-1)}) + \pd_i (\Pi^\perp \dot{\tchi}^{(-1)}) \right] D^{(2)ij} P_j + \cr
&~ + \pd_M \pd_N (\tchi^{(-1)} + \tchi^{(0)}) D^{MP} D^{NQ}\times \cr
&\qquad\times \left( P_P - \smallint \Pi^\perp \dot{\tchi}^{(-1)} \cdot \pd_P \tchi^{(-1)} \right) \left( P_Q - \smallint \Pi^\perp \dot{\tchi}^{(-1)} \cdot \pd_Q \tchi^{(-1)} \right) + \cr
&~ - \pd_{\bx}^2 (\tchi^{(-1)} + \tchi^{(0)}) + g^{-1}
V'(g(\tchi^{(-1)} + \tchi^{(0)}))  + g\; s- \mu^M \pd_M \phi + O(g)~.
\end{align}
We can organize and solve this order by order in the coupling. At
leading order we recover the expression
\begin{align}\label{backgroundeom}
E_{\chi}^{(-1)} =&~ (\pd_i \pd_j \tchi^{(-1)}) D^{(2)ik} D^{(2)jl} P_k P_l - \pd_{\bx}^2 \tchi^{(-1)} + g^{-1} V(g\tchi^{(-1)}) - \mu^{(-1)M} \pd_M \phi~.
\end{align}
Setting $\mu^{(-1)M} = 0$  leads to the
$D$-dimensional generalization of the similar equation in
\cite{Gervais:1975pa}, the solution to which is given by the boosted
soliton profile
\begin{equation}\label{chisol1}
\tilde{\chi}^{(-1)} = g^{-1} \phi\left( \Lambda^{i}_{\phantom{i}j} (x^j - X^j); U^m \right)~,
\end{equation}
where $\phi(\bx - \bX;U^m)$ is the static soliton solution and
\begin{equation}\label{chisol2}
{\Lambda^i}_j = {\delta^i}_j + \left( \sqrt{1 + \frac{\bP^2}{M_{\rm cl}^2}} - 1 \right) \frac{P^i P_j}{\bP^2} ~.
\end{equation}
Regarding the equation of motion for the leading-order Lagrange
multiplier $\mu^{(-1)M}$, we assume that $\chi$, \eqref{B11} with
\eqref{chisol1}, satisfies the orthogonality constraint $\int \chi
\cdot \pd_M \phi = 0$ to leading order. The case of the kink in
$\Phi^4$ theory is consistent with this condition. If the
orthogonality condition on $\chi$ does not hold, then one should
reinstate a nonzero $\mu^{(-1)M}$ and solve the coupled equations.

By further manipulating the $O(1)$ terms in \eqref{Echihybrid} we can
write the equation at this order in terms of a
(linearized)  differential operator
\begin{align}
\mathcal{L} [\tchi^{(0)},\mu^{(0)M}] :=&~ - (\pd_i \pd_j \tchi^{(0)}) D^{(2)ik} D^{(2)jl} P_k P_l + 2 (\pd_i \pd_M \tchi^{(-1)}) D^{(2)ij} D^{(3)Mk} P_j P_k + \cr
&~  - \pd_{\bx}^2 \tchi^{(0)} + V''(g\tchi^{(-1)}) \tchi^{(0)} -
\mu^{(0)M} \pd_M \phi~  \raisetag{16pt}
\end{align}
and the source term
\begin{align}
\mathcal{S}[\tchi^{(-1)}] := &~ - \left[ \pd_t (\Pi^\perp \pd_i
  \tchi^{(-1)}) + \pd_i (\Pi^\perp \dot{\tchi}^{(-1)}) \right]
D^{(2)ij} P_j - g\; s\cr
&  +2 (\pd_i \pd_j \tchi^{(-1)}) D^{(2)ik} D^{(2)jl} P_k \int \Pi^\perp \dot{\tchi}^{(-1)} \cdot \pd_l \tchi^{(-1)} + \cr
&~  - 2 (\pd_i \pd_m \tchi^{(-1)}) D^{(2)ij} D^{(2)mn} P_j \left( P_n
  - \smallint \Pi^\perp \dot{\tchi}^{(-1)} \cdot \pd_n \tchi^{(-1)}
\right) ~, 
\end{align}
such that 
\begin{equation}
E_{\chi}^{(0)} = \mathcal{L} [\tchi^{(0)},\mu^{(0)M}] - \mathcal{S}[\tchi^{(-1)}] ~.
\end{equation}
This can be formally solved by taking
\begin{equation}
(\tchi^{(0)}, \mu^{(0)M} ) = \mathcal{L}^{-1} \mathcal{S}[\tchi^{(-1)}]~.
\end{equation}

\bibliographystyle{utphys}
\bibliography{monopole}

\end{document}